\documentclass[sigconf,screen]{acmart}

\AtBeginDocument{%
  }

\setcopyright{none}
\acmConference[]{}{}{}
\renewcommand\footnotetextcopyrightpermission[1]{} 
\usepackage{graphicx}
\usepackage{algorithmic}
\usepackage{textcomp}
\usepackage{xcolor}
\usepackage{paralist}
\usepackage{xspace}
\usepackage{booktabs}
\usepackage{hyperref}
\usepackage{siunitx}
\usepackage{subcaption}
\usepackage[skip=0.5\baselineskip]{caption}

\usepackage{tabularx}
\usepackage{svg}

\usepackage{tcolorbox}
\usepackage{comment}
\usepackage[ruled,vlined]{algorithm2e}
\usepackage[inline]{enumitem}
\usepackage{listings}

\makeatletter
\def\lst@makecaption{%
  \def\@captype{figure}%
  \@makecaption
}

\lstdefinelanguage{CSharp}{
  language=[Sharp]C,
  morekeywords={async,await,var}
}

\usepackage{cleveref}

\newtcolorbox{thebox}[1][]{
  title    = {#1},
  #1,
  top=1mm,
  bottom=1mm,
  left=1mm,
  right=1mm,
  toptitle=0mm,
  bottomtitle=0mm,
  boxsep=0.5mm
}

\newcommand{\ie}{\emph{i.e.,}\xspace}
\newcommand{\eg}{\emph{e.g.,}\xspace}

\newcommand{\etal}{\emph{et al.}\xspace}
\newcommand{\etc}{\emph{etc.}\xspace}
\newcommand{\UltraInstinctVR}{\textsc{Ul\-tra\-In\-stin\-ct\-VR}\xspace}
\newcommand{\VRGreed}{\textsc{VR\-Gre\-ed}\xspace}
\newcommand{\VRGuide}{\textsc{VR\-Gui\-de}\xspace}
\newcommand{\VRExplorer}{\textsc{VR\-Explo\-rer}\xspace}
\newcommand{\XRintTest}{\textsc{XR\-int\-Test}\xspace}
    
\begin{document}

\title{System Test Generation for Virtual Reality Applications using Scenario Models
}

\author{Gerry Longfils}
\orcid{0009-0005-3286-0720}
\email{gerry.longfils@unamur.be}
\author{Maxime Cauz}
\orcid{0000-0002-1234-1772}
\email{maxime.cauz@unamur.be}
\affiliation{%
  \institution{NADI, University of Namur}
  \city{Namur}
  \country{Belgium}
}

\author{Arnaud Blouin}
\orcid{0000-0002-5974-9601}
\affiliation{%
  \institution{University of Rennes, INSA Rennes, IRISA, Inria}
  \city{Rennes}
  \country{France}
}
\email{arnaud.blouin@irisa.fr}

\author{Xavier Devroey}
\orcid{0000-0002-0831-7606}
\affiliation{%
  \institution{NADI, University of Namur}
  \city{Namur}
  \country{Belgium}
}
\email{xavier.devroey@unamur.be}

\begin{abstract}
Virtual Reality (VR) applications are increasingly being integrated across a wide range of domains, including surgical training and industrial marketing. However, the long-term adoption and maintenance of VR applications remain limited, particularly due to the lack of effective, systematic, and reproducible software testing approaches tailored to their unique characteristics.
To address this issue, we introduce \UltraInstinctVR, a novel testing approach for VR applications. Relying on predefined VR models (scenarios), it automates the generation and execution of concrete VR system tests. In our empirical evaluation, we compare \UltraInstinctVR with state-of-the-art automated VR testing approaches in terms of coverage and failure detection on 10 open-source VR applications. The results show that \UltraInstinctVR outperforms existing automated tools for detecting unique failures and provides valuable insights for identifying real-world bugs in VR applications.
\end{abstract}
\keywords{Virtual reality, Interaction Testing, Model-based Testing, System Testing}

\maketitle

\section{Introduction}

%
Virtual Reality (VR) is an emerging technology that enables users to engage with immersive, computer-generated worlds~\cite{agrawal2019defining}, typically through a Head-Mounted Display (HMD) \cite{sutherland1965ultimate}.
During the last decade, the development of VR applications has grown thanks to the availability of dedicated hardware devices (\eg HMDs) and the distribution of VR applications on various application stores. 
However, the software engineering practices and tools supporting the development of such applications (\eg available standard frameworks, integrated development environments, dependency management, automated tools, \etc) still have a long way to go.

%
The immersive and highly interactive characteristics of VR applications introduce new software engineering challenges, particularly in software testing, beyond those encountered in software and classical graphical user interface (GUI) testing~\cite{rzig2023virtual, ashtari_creating_2020}. First, VR applications tend to involve more natural user interactions (\eg gestures, voice, and gaze) moving away from more traditional interactive devices, such as keyboard and mouse. 
Second, the development of VR applications relies on more complex mathematical interactions, detection, and rendering processes that involve the inclusion of the third spatial dimension, just as in video game development.
Third, the faults that affect 3D environments~\cite{rzig2023virtual} and VR applications~\cite{andrade2019towards} extend those of classical 2D user interfaces~\cite{lelli2015classifying} to include additional human-based interactions, input devices, and performance-related faults.

The current state of the art includes a clear understanding of the various fault categories existing in VR~\cite{andrade2020understanding}. However, few testing approaches have been designed to detect them efficiently.
Existing VR testing approaches are limited to crash detection \cite{wang2022vrtest, zhu2025vrexplorer, gu2025xrinttest}, which limits their usefulness to efficiently testing VR applications.
Integrating other kinds of VR and 3D faults within VR testing approaches is challenging.
First, test assertions follow complex user interaction sequences executed in complex input spaces (\ie 3D environments). 
Second, such input spaces are very large, as the user can engage in various interactions at any time, leading to a combinatorial explosion of the number of cases to test.
As a consequence, current VR testing practices remain mainly manual~\cite{rzig2023virtual, ashtari_creating_2020} or do not consider specific VR and 3D faults~\cite{wang2022vrtest}.

%
This paper proposes a first step towards a more mature and automated software testing process for VR applications: \UltraInstinctVR, a novel Model-Based Testing (MBT) approach to automate the \textit{generation} and \textit{execution} of VR system tests.
The generated tests detect failures related to widely-used VR \textit{user interactions}, namely: \textit{selection}, \textit{locomotion}, and \textit{manipulation tasks}~\cite{spittle2022review}.
Our proposal builds on recent VR development approaches that define different models during the development process~\cite{lemoulec:hal-01613873, claude2014short}.
In particular, our approach leverages VR scenario models that describe sequences of interactions within VR applications and are used to automate tasks such as moving or interacting with virtual objects. For instance, such technologies can be used to create demonstration videos in a 3D environment. 

Such scenario models can take the form of Petri nets and are usually used to guide users at run time in performing a task~\cite{claude2014short}.
A tester then configures a test campaign by mapping predefined assertions serving as an oracle mechanism to scenarios, and by specifying user interactions' parameters (\eg, size of the collider required for selection). 
Based on those inputs, \UltraInstinctVR generates multiple concrete interaction sequences to build test cases that can be easily re-executed. 

Unlike prior approaches that primarily focus on automatically generating interactions, our Petri net-based method combines user interactions sequencing with assertions. This enables the automated generation of functional tests. Our implementation further builds upon a user interaction taxonomy \citep{spittle2022review} established by the VR community, thereby contributing to the automation of functional VR testing and, more broadly, to improving the overall quality of VR software.


To evaluate \UltraInstinctVR, we developed a prototype using Unity, a well-known 3D engine, and Xareus \cite{Xareus}, a VR scenario engine plugin for Unity that allows the definition of informative interaction scenarios.
We conducted a three-phase evaluation to:
\begin{enumerate}
    \item compare \UltraInstinctVR's ability to generate system tests able to cover lines of code and interactions with virtual objects (\textbf{RQ1}) against state-of-the-art \VRGreed and \VRGuide approaches, based on search algorithms to identify interactive objects; 
    \VRExplorer, a model-based testing tool designed to interact with diverse virtual objects and explore complex VR scenes;
    and \XRintTest, an automated testing framework for generating interactions in Unity-based XR applications. 
    \item analyze the types of test failures triggered by the different tools (\textbf{RQ2}) across ten open-source VR applications.
    \item assess the capability of \UltraInstinctVR to detect real-world VR defects (\textbf{RQ3}).
\end{enumerate}
%

The evaluation results show that \UltraInstinctVR outperforms \VRExplorer and \XRintTest in terms of line coverage and interaction coverage. Regarding \VRGreed and \VRGuide, although the coverage mostly shows no statistically significant differences, the diversity of test failures triggered by \UltraInstinctVR is almost doubled compared to \VRGreed and \VRGuide. This underlines the ability of \UltraInstinctVR to cover different behaviors compared to other approaches. Finally, our manual analysis of the different failures confirmed the ability of \UltraInstinctVR to detect real-world VR defects causing such failures.

%
To summarize, this paper makes the following scientific contributions:
\begin{enumerate}
    \item A novel approach \UltraInstinctVR to produce and execute system tests for VR applications;
    \item A prototype developed on a widely used 3D engine, Unity, and a model-based VR plug-in, Xareus;
    \item A quantitative empirical evaluation that demonstrates that the proposed approach outperforms existing automated failure detection tools, particularly in the early stages of execution;
    \item A companion replication package that contains the code of our prototype and details regarding the evaluation.\footnote{\UltraInstinctVR replication package available at \url{ https://doi.org/10.5281/zenodo.
19248284.}}
    \footnote{Code of \UltraInstinctVR \url{https://doi.org/10.5281/zenodo.
19248284.}}
\end{enumerate}

\section{Background and Related Work}\label{sec.related}

\subsection{Virtual Reality Application Testing}

Andrade et al.~\citep{andrade2019towards} identified a lack of structured testing methodologies for VR systems.
They showed that VR projects exhibit a higher degree of fault-proneness compared to non-VR applications. The same authors proposed a taxonomy of eight main defect categories, including physics-related errors, stability problems, and user interaction issues~\citep{andrade2020understanding}.
It highlights the need for specialized testing approaches targeting the specific challenges of VR development. In this context, user interaction-related defects are particularly critical due to their impact on user experience and performance.

The primary approach to VR testing remains manual testing~\citep{gu2025software}, where testers interact with the VR environment using a headset to validate the behavior of the application. However, manual testing suffers from limited scalability and reproducibility, making it unsuitable for large-scale testing efforts. To address this, automated approaches such as \textit{VRTest}~\citep{wang2022vrtest} simulate virtual users navigating VR scenes using raycasting to detect interactive objects. VRTest integrates multiple strategies, including \VRGreed, which employs a greedy exploration algorithm, and VRSmart, which combines greedy and random exploration to increase the scene coverage. \VRGuide~\citep{wang2023vrguide} extends VRTest with dynamic cut coverage for more efficient path exploration, but like VRTest, it only detects \textit{NullReferencesExceptions}.

Other approaches leverage haptic interfaces, such as the record-and-playback technique proposed by Corrêa et al.~\citep{correa2021software}, which captures haptic inputs and applies inference rules to detect object collisions. This approach, however, faces broader limitations due to bidirectional device communication, fluctuating frame-update rates, the difficulty of reproducing expert actions, and the subjective nature of validating haptic feedback. Similarly, PLUME~\citep{10458415} supports the recording, playback, and analysis of VR interactions but is limited to single-user recordings and still requires manual analysis, as testers must explicitly record and replay interaction traces. In contrast, our approach aims to automate the testing process as much as possible, reducing the reliance on manual intervention.

Automated exploration techniques, which run the software under test at runtime and explore scenes in both 2D and 3D games, require robust auto-navigation and exploration capabilities, as built-in game navigation algorithms often do not provide sufficient control for testing purposes~\citep{prasetya2020navigation}. The iv4XR Framework offers an agent-based approach to automate interactive system testing.
It currently targets 3D games, with future work planned for VR systems, integration of learning-based AI, and broader evaluation across pilot studies~\citep{prasetya2021agent}.

Compared to \textit{VRTest}, the \textit{iv4XR} framework adopts a fundamentally different exploration and testing strategy. iv4XR relies on graph-based representations of the environment and advanced reasoning over agent behaviors. In contrast, \textit{VRTest} primarily employs random navigation combined with greedy heuristics to explore VR scenes, which may lead to less systematic coverage. Moreover, while iv4XR provides a general-purpose framework for testing interactive and agent-based systems, it has not yet been adapted and evaluated for VR system-level testing, whereas \textit{VRTest} specifically targets VR applications.

More recently, researchers have shown an increasing interest in interaction-based testing for XR applications. Gu et al. developed \textit{INTENXION}~\citep{gu2025test}, a test automation approach for XR applications implemented in Unity. The framework leverages a taxonomy of XR interactions to systematically cover tasks such as navigation, object manipulation, and selection. Their case study demonstrates that INTENXION can effectively automate a wide range of XR user interactions.
In subsequent work, the same authors introduced \textit{XRIntTest} \citep{gu2025xrinttest}, an approach for automating user interaction testing in XR applications developed in Unity. This approach addresses limitations of existing XR testing methods, particularly their difficulty in simulating realistic 3D spatial interactions produced by 6-DoF controllers. XRIntTest constructs an interaction graph that models required interaction events (\eg \textit{selection}, \textit{grab}) and uses it to automatically explore scenes and generate user actions. Evaluated on the XRBench3D benchmark, the framework achieves high coverage of interaction on virtual objects and proves significantly more effective and efficient than random testing, while also detecting runtime exceptions.

Finally, Zhu et al.~\citep{zhu2025vrexplorer} present VRExplorer, an automated testing approach for Unity-based VR applications that performs systematic scene exploration and interaction with virtual objects. The approach introduces the \textit{Entity–Action–Task} (EAT) hierarchical framework to generically model VR interaction behaviors across different Unity versions and interaction plugins. Built on this model, the VRExplorer agent combines NavMesh-based pathfinding for autonomous navigation with a probabilistic finite-state machine to execute diverse interactions. An extensive evaluation on 11 VR projects shows that VRExplorer significantly outperforms the state-of-the-art VRGuide in both code and method coverage. Additionally, the approach successfully detects previously unknown functional and non-functional bugs, demonstrating its effectiveness for automated VR testing.

However, the main limitation of these approaches is their primarily focus on crash detection and limited support for identifying other categories of faults.

Ultimately, exploratory studies of 320 open-source VR projects on GitHub and Stackoverflow reveal that most VR projects are small-scale, single-developer efforts with limited collaboration, few releases, and minimal use of version control features~\citep{ghrairi2018state}. Despite the widespread use of Unity and popular languages such as C\#, VR software testing remains immature and largely reliant on manual testing~\citep{andrade2019towards}, which is often non-reproducible, non-automated, and unreliable. This situation highlights the need for specialized tools, techniques, and structured interaction testing approaches to better support software quality.

\subsection{Model-based and Graphical User Interface Testing}
\label{sec:mbt}
%

MBT techniques rely on models to automatically generate tests~\citep{utting_practical_2007}. 
However, the major limitation of MBT techniques is the significant cost associated with creating models specifically for testing purposes~\citep{10.1145/3715789}. To mitigate this issue in traditional software domains, several GUI testing approaches for desktop, web, and mobile systems have introduced run-time model extraction techniques commonly referred to as GUI ripping~\citep{gu2019practical,10.1145/2351676.2351717}. These techniques automatically infer models representing GUI layouts, basic events (\eg mouse clicks), and possible execution paths, often encoded as state machines, event-flow graphs, or Markov chains~\citep{6786194,lelli:hal-01123647,6823874}.
Additionally, these models are recommended to be designed early in the application development process in the context of model-based GUI developments~\citep{calvary2014introduction}. Such models can later be reused for MBT. This practice supports the translation of user requirements into UI implementations. It is highly recommended to define such models, as they improve communication between teams while ensuring traceability and system maintainability. However, these approaches primarily target 2D interfaces composed of widgets such as buttons or menus.

Other model-based approaches for software implementation have been developed to facilitate the development of VR applications. Among these, scenario engines such as \#SEVEN~\citep{claude2014short} support the specification of temporal constraints, concurrency, and user-driven interactions. These engines are built upon formal models such as Petri nets or state machines, which developers traditionally use to define interaction scenarios that guide end users through immersive environments. 
Beyond their original purpose in VR development, these formal models can also be leveraged for MBT. In an MBT context, the same behavioral models used to define user scenarios can serve as test models, from which interaction sequences and expected outcomes are automatically derived. This dual use enables systematic test case generation and oracle specification based on formally defined behaviors. 
Xareus~\citep{Xareus,gramoli2025xareus} exemplifies this approach. It relies on expressive formal models, notably Petri nets, to structure and execute task sequences in immersive environments. Xareus combines a scenario engine for behavioral specification, a relation engine to manage dependencies among virtual objects, and a collaborative interaction component to support multi-user coordination. While such expressiveness enables the modeling of complex and collaborative scenarios, it also introduces additional modeling effort.

Testing VR applications raises additional challenges due to the complexity of human-centered 3D interactions. 
According to Spittle et al.~\citep{spittle2022review}, VR applications span multiple interaction modalities, including \textit{freehand}, \textit{speech-based}, \textit{head-based}, and \textit{hardware-based} interactions. 
This diversity substantially expands the space of possible test cases and limits the applicability of traditional testing techniques (\eg unit testing). In this context, MBT provides a promising solution, as it can systematically handle the large and multidimensional interaction space inherent to VR applications and provide a real alternative to \textit{manual testing}.

\section{The \UltraInstinctVR approach}

\begin{figure*}[t]
    \centering
    \subfloat[Steps of the approach]{%
        \includegraphics[width=84mm]{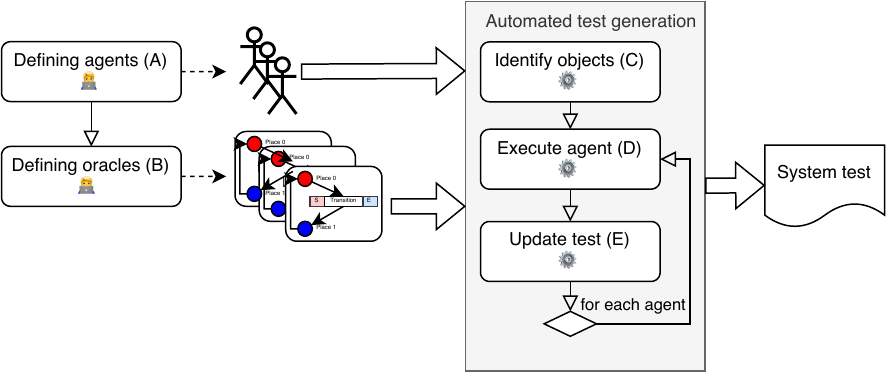}%
        \label{fig:ultrainstinct:steps}%
    }%
    \hfil%
    \subfloat[Example of a test agent performing simple teleportation with its associated Petri net serving as oracle for a given VR scene]{%
        \includegraphics[width=84mm]{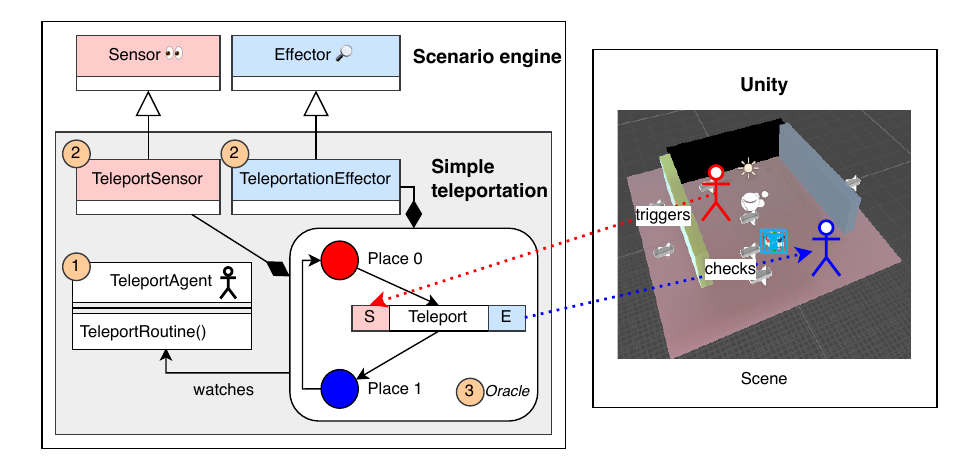}%
        \label{fig:ultrainstinct:approach}%
    }%
    \caption{\UltraInstinctVR testing approach} 
    \label{fig:ultrainstinct}
\end{figure*}

We introduce \UltraInstinctVR, a user interaction testing approach that automatically generates VR system tests. \UltraInstinctVR is a model-based testing approach that leverages VR scenario models that describe sequences of expected interactions to be executed by users within VR applications.

In our approach, we consider scenario models~\citep{claude2014short} to define an \textit{oracle} for VR applications. This oracle is represented as a Petri net assigned to a test agent, which performs a sequence of interactions in the VR environment. In this case, a test agent is considered a (non-AI) automated player that performs actions within the scene during the runtime execution of the software under test. The Petri net then verifies whether the actions executed in the VR application are correct from the specification point of view.

The different steps of our approach are detailed in \Cref{fig:ultrainstinct:steps}.
First, the tester manually defines test agents (A), which are responsible for executing actions within the scene (\ie the VR environment, as illustrated on the right side of \Cref{fig:ultrainstinct:approach}). Testers can define new agents, each exhibiting different behaviors tailored to their specific needs (\eg (1) in \Cref{fig:ultrainstinct:approach}). For instance, an agent can perform interactions such as teleporting outside the scene and grabbing objects. Second, the tester has to associate each agent with a Petri net that describes valid interactions~(B).
This Petri net will serve as an \textit{oracle} so that, when specific interactions or sequences of interactions are performed, assertions verify that the results are correct.
For that, each transition of a Petri net has: 
(i) a \textit{Sensor} that acts as a trigger detecting a specific interaction (\eg teleportation); and 
(ii) an \textit{Effector} that performs checks on the effects of the interaction (\eg checking that the agent is still in the scene's bounds). See, for instance, (2) and (3) in \Cref{fig:ultrainstinct:approach}.

Then, automated test generation begins with an analysis to identify interactive virtual objects in the scene (C). After this step, for each test agent, \UltraInstinctVR executes the agent's interactions (\eg teleportation in \Cref{fig:ultrainstinct:approach}) and, for each interaction, checks whether a transition has been triggered in the associated Petri net (D) to let the corresponding Effector check the new state of the scene (denoted in blue in \Cref{fig:ultrainstinct:approach}). If an effector detects a violation, it reports the error to the tester along with the corresponding failing test. Finally, the agents' behaviors are recorded sequentially to produce an automated system test (E).

\subsection{Test agents definition}
\label{sec:agents_def}

In our approach, each test agent is assigned a specific task within the VR scene, namely, triggering the actions under test. To accomplish these tasks, agents interact with objects and elements in the virtual environment. Conceptually, different agents are responsible for exercising different categories of interactions. For example, one agent performs locomotion by navigating to various points in the VR scene, while another agent manipulates virtual objects. In practice, our implementation provides a set of predefined agents listed in \Cref{tab:test-categories}. These agents can be reused by activating them during the test generation process. For instance, one agent performs valid teleportation (\textit{simple teleportation}), while two others attempt invalid teleportation scenarios (\textit{teleportation outside scene bounds} and \textit{teleportation into objects}). Additional test agents can be defined depending on the testing requirements.

\Cref{alg:interaction-generation} defines the generation of the interaction. First, before executing the interaction, the tester must set up the parameters that define the interaction's precision and the other parameters that influence it. For instance, the hand size used to grab the virtual object, and the precision with which the hand interacts with it. 
The set of interactions to generate (\textit{interactionToPerform}) and the delay in seconds between two interactions are provided as inputs. The agent then enters a loop in which, at each iteration, it performs an interaction from the interaction set assigned to it. The interaction is selected based on the execution sequence of interactions, as parameterized by the tester.  and applies the corresponding parameters specified by the tester to the test agent. The interaction is then executed in the virtual environment, after which the agent waits for the specified delay before generating the next interaction.

\begin{algorithm}[t]
\caption{Generic interaction generation by a test agent}
\label{alg:interaction-generation}
\KwIn{ $iterationToPerform$, $delay$}
\ForEach{$interaction \in iterationToPerform$}{
    $params \leftarrow$ GenerateInteractionParameters($interaction$)\;
    ExecuteInteraction($interaction$, $params$)\;
    Wait($delay$)\;
}
\end{algorithm}

\subsection{Defining oracles}

Our approach relies on the scenario engine proposed by Claude \etal~\citep{claude2014short} to define test oracles. As illustrated in (3) in \Cref{fig:ultrainstinct:approach}, the oracle is represented as a circular Petri net. For each transition, the tester specifies the corresponding \textit{Sensor} and \textit{Effector} (\eg (2) in \Cref{fig:ultrainstinct:approach}). The sensor is responsible for detecting relevant interactions (\ie triggering the transition), while the effector verifies their effects. The same sensor or effector may be reused across multiple transitions and Petri nets.
More specifically, the sensors automatically detect events generated by the test agent. When an event is detected (\eg the selection of a virtual object), the corresponding transition in the Petri net is fired, which activates the associated effector. The effector then verifies whether the interaction has been correctly executed by the test agent.

To illustrate conceptually, the sensor defined in Algorithm~\ref{alg:sensor} detects interactions performed by the test agent and can be reused across the Petri nets associated with different agents. The sensor observes the current state of the environment and compares it with the previously recorded state to determine whether a known interaction has occurred. For each interaction type, a detection condition evaluates whether the observed state change corresponds to that interaction. When such a condition is satisfied, the sensor creates an interaction event containing the relevant contextual information and triggers the corresponding Petri net transition by returning \textit{true}.

\begin{algorithm}[t]
\caption{Generic sensor for interaction detection}
\label{alg:sensor}

\KwIn{$interactionToPerform$, $previousState$}

$currentState \leftarrow$ ObserveEnvironment()\;

\ForEach{$interaction \in interactionToPerform$}{
    \If{DetectInteraction($interaction$, $previousState$, $currentState$)}{
        $event \leftarrow$ CreateInteractionEvent($interaction$, $currentState$)\;
        TriggerTransition($event$)\;
        \Return $(true, event)$\;
    }
}

$previousState \leftarrow currentState$\;
\Return $(false, \varnothing)$\;

\end{algorithm}

\begin{algorithm}[t]
\caption{Generic effector for oracle verification}
\label{alg:effector}

\KwIn{$oracleSet$}

$interaction \leftarrow$ SafeEffectorUpdate()\;
$oracle \leftarrow$ RetrieveOracle($interaction$, $oracleSet$)\;

$currentState \leftarrow$ ObserveEnvironment()\;

\If{Verify($oracle$, $event$, $currentState$)}{
    ReportSuccess($interaction$)\;
}
\Else{
    ReportFailure($interaction$)\;
}

\end{algorithm}

\Cref{alg:effector} presents the conceptual behavior of the effector. The \texttt{SafeEffectorUpdate} method provided by the scenario engine is overridden to process interaction events detected by the sensor. For each interaction, the effector retrieves the corresponding oracle, observes the current system state, and verifies whether the expected condition is satisfied. It then reports either a successful interaction or a failure, depending on the verification result.

Separating the agent's behavior (\eg teleporting in \Cref{fig:ultrainstinct:approach}) from the oracle checking the result of an interaction allows decoupling the \textit{action} from the \textit{assertion}, following the recommended \textit{Arrange-Act-Assert} testing pattern \cite{aniche2022effective}.

\subsection{Identify interactive objects in the scene}

For detecting interactive objects in the scene (step (C) in \Cref{fig:ultrainstinct:steps}), we adapt the \VRGreed algorithm \citep{wang2022vrtest}. Specifically, we compute the number of newly detected interactive objects per minute. We define an interactive object as any object that a user can interact with. Once this value drops to zero, the scanning process terminates, and the algorithm stops executing. Each detected object is stored along with its position, providing the test agents with a database of objects to interact with during testing.
This process also detects failures, such as \texttt{NullReferencesExceptions} or \texttt{ObjectNotFoundReferences} caused, for instance, by a missing or ill-defined parameter value in the detected object.

\subsection{Execution of the test agent}

During test generation, the different agents are initialized and operate in the VR scene one after the other (steps (D) and (E) in \Cref{fig:ultrainstinct:steps}). \Cref{fig:ultrainstinct:approach} provides an example of test generation for a single agent performing \textit{simple teleportation} in valid places. The associated Petri net, which checks that teleportation is effectively performed, is initialized in \textit{Place 0}.
In the example in \Cref{fig:ultrainstinct:approach} The agent will then start interacting within the scene according to the routine defined in \texttt{TeleportRoutine()}. After each interaction, the sensor, described in \Cref{alg:sensor}, checks if the expected interaction event is detected and triggers (or not) the corresponding transition, which activates the effector described in \Cref{alg:effector}. The execution is repeated for each defined agent.

Test agents can operate with a controlled degree of variability when interacting within the VR environment. Testers can configure parameters that influence how interactions are executed. For example, \textit{teleportation} includes setting the \textit{number of teleportation attempts}, which determines how many times the agent triggers the action, and the \textit{teleportation threshold}, which specifies the minimum distance required for a movement to be considered a teleportation. For \textit{object grabbing} and \textit{collision} in \Cref{tab:test-categories}, testers may adjust the \textit{collider sphere radius}, which controls the sensitivity and precision of object detection: a larger radius allows less precise interactions regarding object positioning, whereas a smaller radius enforces strict accuracy.

\subsection{Update of the system test}

Test agents are executed sequentially, each operating independently to evaluate a specific aspect of user interaction in the VR scene. This structure organizes the testing process in a controlled, systematic way. By separating interactions into individual agents, the approach provides clearer analysis and improves isolation during execution, guaranteeing deterministic test case generation. For interactions involving nondeterministic behavior, such as teleportation to arbitrary locations within or outside the scene, we record the exact execution sequence and parameter values used for the interaction (\eg destination location). This allows testers to deterministically replay the same scenario, thereby reducing flaky tests~\citep{parry2021survey}, a limitation often observed in existing approaches.

The sequence in which agents are executed is left to the tester to prevent undesired side effects. In particular, agents that rely on objects position (\eg valid teleportation in our case) should be executed before agents modifying the scene (\eg agents moving objects). An alternative could be to reinitialize the scene between the execution of two agents. However, in practice, such an approach has been shown to slow down the generation process. As shown in our example, isolating agents is still possible using \UltraInstinctVR, with each agent used separately.

\section{Implementation}
\label{sec:implem}

\begin{table}[t]
\centering
\footnotesize
\caption{List of agents' behaviors and their corresponding oracles currently available in \UltraInstinctVR}
\label{tab:test-categories}
\begin{tabularx}{\columnwidth}{p{20mm}p{13mm}X}
\toprule
\textbf{Agent's behavior} & \textbf{Interaction family~\citep{spittle2022review}} & \textbf{Description of the oracle} \\
\midrule
\textbf{Simple Teleportation} & Use Case & Verifies that the user can teleport to valid locations within the virtual environment. The teleportation sequences are performed randomly. \\
\textbf{Teleportation Outside Scene Bounds} & Use Case & Assesses how the system handles attempts to teleport beyond the scene's predefined boundaries. Teleportations are also performed randomly. \\
\textbf{Teleportation Into Objects} & Use Case & Tests the application behavior when teleporting into the physical space occupied by scene objects. \\
\textbf{Object Selection} & Tasks & Ensures that all interactive objects in the environment can be correctly selected through standard user input methods. \\
\textbf{Object Grabbing} & Tasks & Validates that objects can be grabbed using VR controllers. \\
\textbf{Object Movement} & Tasks & Confirms that once grabbed, objects can be repositioned, particularly to a designated central location in the scene. \\
\textbf{Collision} & Tasks & Confirms that objects can collide with each other appropriately. \\
\bottomrule
\end{tabularx}
\end{table}

\UltraInstinctVR is implemented using \textit{Unity} and \textit{Xareus} \citep{gramoli2025xareus}, a scenario-based framework that implements the approach proposed by Claude \etal~\citep{claude2014short}. 
We provide several default test agents and their corresponding oracles, listed in \Cref{tab:test-categories}, each one belonging to an interaction family defined by Spittle et al.~\citep{spittle2022review}. The test agents were designed to be as generic as possible, enabling their reuse across a wide range of VR applications, from simple navigation within a virtual scene to more complex interactions involving the manipulation of multiple objects. The complete implementation is open source.

Our current implementation supports three fundamental categories of user interactions commonly found in VR applications: \textit{locomotion}~\citep{lee2023comparison}, \textit{selection}~\citep{atienza2016interaction}, and \textit{manipulation}. Locomotion refers to navigation within the virtual environment, such as natural walking or point-and-click teleportation. Selection involves identifying and choosing virtual objects, while manipulation includes actions such as grabbing, rotating, and repositioning objects. These interaction types align with two of the four interaction families identified by Spittle et al.~\citep{spittle2022review}, namely \textit{use case} and \textit{task}.

The agents and oracles provided by \UltraInstinctVR are designed to be reusable across a wide range of VR applications. To maximize generality, we implemented common interaction scenarios such as simple teleportation, object grabbing, and object movement within virtual space. Testers can further adapt and refine these models to meet application-specific requirements, provided they understand the underlying \textit{Sensor} and \textit{Effector} mechanisms.


Currently, our implementation does not cover less common VR interaction modalities, such as voice-based interactions. Supporting such interaction families would not affect the proposed approach, but would require additional development effort. In particular, the two remaining interaction families identified by Spittle et al. \textit{study type} and \textit{input methods} are left for future work, as they involve advanced interaction mechanisms (\eg voice commands or eye-tracking) that are more difficult to model and automate~\citep{10.1145/3660515.3664244}.

\section{Empirical evaluation}


Our empirical evaluation focuses on the following research questions:
\begin{compactitem}
\item[\textbf{RQ1}] How does \UltraInstinctVR perform in terms of coverage compared to state-of-the-art automated VR testing approaches?
    \begin{compactitem}
        \item[\textbf{RQ1.1}]How does \UltraInstinctVR compare to existing automated VR testing tools in terms of line coverage? 
        \item[\textbf{RQ1.2}]How does \UltraInstinctVR compare to existing automated VR testing tools in terms of interactive virtual object coverage?
    \end{compactitem}
    \item[\textbf{RQ2}] Which categories of runtime failure are revealed by \UltraInstinctVR during automated VR benchmark executions?
\item[\textbf{RQ3}] To what extent \UltraInstinctVR can detect real-world defects?
\end{compactitem}

\textbf{RQ1} investigates the ability of automated VR testing approaches to execute the different parts of the applications. For that, we rely on standard line coverage (\textbf{RQ1.1}), which indicates which parts of the source code have been executed. We also rely on interaction coverage (\textbf{RQ1.2}), a metric used in VR to measure the percentage of objects in the VR scene that have been interacted with (\eg grabbed or moved). We then investigate in \textbf{RQ2} the type of failures (\ie failing test executions) that are triggered by the different approaches (\eg assertion violation or application crash), and to what extent such failing test executions can help fix underlying defects in \textbf{RQ3}.

\subsection{Evaluation setup}

\begin{table*}[t]
    \centering
    \caption{Selected Benchmarks with the GitHub repository, the number of files, the total number of lines of code (LOC), game objects, and Unity version.}
    \label{tab:benchmarks}
    \small
    \begin{tabular}{l l r r r r r}
    \toprule
        \textbf{Project} & \textbf{GitHub Repository} & \textbf{Files}  & \textbf{LOC} & \textbf{Game Objects} & \textbf{Unity Version} & \textbf{Stars}\\
    \midrule
        BaM-Construct & \href{https://github.com/digitalplusplus/BaM-Construct}{digitalplusplus/BaM-Construct} & 2444 & 26965 & 362 & 6000.0.23f1 & 3 \\
        LayoffInvader & \href{https://github.com/martin226/layoffevaders}{martin226/layoffevaders} & 2800 & 5306 & 19 & 6000.0.32f1 & 7 \\
        VertexForm3d & \href{https://github.com/Vertex-Form-3D/vertexform3d-unity-vr-starterkit}{Vertex-Form-3D/vertexform3d-unity-vr-starterkit} & 7851 & 11473  & 1202 & 6000.1.3f1 & 154\\
        EscapeRoom & \href{https://github.com/ruizhengu/XRBench3D}{ruizhengu/XRBench3D}  & 40893 & 1761076 & 162 & 6000.0.45f1 & 1\\

        VR-Game-Jam-Template & \href{https://github.com/ruizhengu/XRBench3D}{ruizhengu/XRBench3D} & 36743 & 1546558  & 206 & 6000.0.45f1 & 1\\
        XRI Examples & \href{https://github.com/ruizhengu/XRBench3D}{ruizhengu/XRBench3D}  & 29399 & 1389774 & 862 & 6000.0.45f1 & 1\\
        XRI Starter Assets & \href{https://github.com/ruizhengu/XRBench3D}{ruizhengu/XRBench3D} & 32250 & 1523732  & 192 & 6000.0.45f1 & 1\\
        XRI Starter Kit & \href{https://github.com/ruizhengu/XRBench3D}{ruizhengu/XRBench3D} & 34665 & 15396513  & 171 & 6000.1.3f1 & 1\\
        XRToolKitEssentials & \href{https://github.com/ruizhengu/XRBench3D}{ruizhengu/XRBench3D}  & 29492 & 1418753  & 2944 & 6000.0.45f1 & 1\\
        VRTemplate & \href{https://github.com/ruizhengu/XRBench3D}{ruizhengu/XRBench3D}  & 29492 & 1418753  & 2944 & 6000.0.45f1 & 1\\
    \bottomrule
    \end{tabular}
\end{table*}

\paragraph{Projects Selection}

For our evaluation, we built a dataset of open-source GitHub projects.
We selected the projects based on the following inclusion criteria: 
(i) at least three GitHub stars, indicating a minimal level of community recognition, to ensure the projects are representative (except for \textit{VR Unity6 Template}, which is the official VR template provided by Unity and the benchmark project proposed by \textit{XRBench3D}~\citep{gu2025xrinttest});
(ii) a last commit performed within the past three years to ensure the project is still maintained;
(iii) the usage of the OpenXR plugin to ensure compatibility with \UltraInstinctVR's requirements; 
(iv) a compatibility with Unity version 6.0 or later to avoid compatibility issues.
We used the GitHub REST API~\citep{GithubAPI} through a custom Python script to retrieve and filter repositories. 

This process yielded 20 candidate repositories. We then performed a manual validation step to exclude false positives, such as repositories that referenced OpenXR in their manifest files without actually using it, or those that did not implement any VR-specific features. We obtained a final set of four repositories (see \Cref{tab:benchmarks}). Each of them compiles and implements representative VR user interactions, such as hand or controller manipulation, which are relevant to evaluate \UltraInstinctVR.
The selected repositories show no evidence of automated tests (\eg unit tests), which can be attributed to the predominance of manual testing in VR.
We also reuse projects from XRBench3D~\citep{gu2025xrinttest}, bringing the total to 10 projects. 
We cannot include the projects used by VRExplorer~\citep{zhu2025vrexplorer}, as they rely on older versions of Unity (2022.X and earlier), which are incompatible with our implementation. As mentioned, the implementation of \UltraInstinctVR relies on dependencies such as Xareus~\citep{gramoli2025xareus}, which are no longer supported in recent versions of Unity.

\paragraph{Data collection}

For data collection, we recorded the number of detected failures. We launched each testing approach for 20 minutes. Note that \UltraInstinctVR may terminate earlier due to predefined stopping conditions (\ie when all system tests have been completed). In contrast, the other approaches run for a fixed duration of 20 minutes, as they do not implement explicit termination criteria. Their execution only stops prematurely in the event of a crash, such as a \textit{NullReferenceException}, as dictated by the Unity Test Framework. This duration was selected based on the hypothesis that \VRGreed and \VRGuide would complete the analysis of the entire VR scene in this time.
Since \VRExplorer and \XRintTest default to a 10-minute testing time budget, we used the same time limit when executing their approaches.

All experiments were conducted on a desktop workstation equipped with an Intel Core i9-10900C processor (3.70~GHz), 32~GB of RAM, a 1.86~TB SSD, and an NVIDIA GeForce RTX 3080, running Windows~10. Different Unity versions were used depending on project requirements.

\emph{Test case generation and coverage analysis (RQ1)}.
We conducted a benchmark comparison against five automated tools for system-level VR testing: \VRGreed \citep{wang2022vrtest}, \VRGuide \citep{wang2023vrguide}, \VRExplorer \citep{zhu2025vrexplorer}, and \XRintTest \citep{gu2025xrinttest}. \VRGreed \citep{wang2023vrguide} employs a greedy exploration algorithm to explore the VR scene and identify interactive elements, triggering runtime exceptions.

\VRGuide also extends the approach by introducing the dynamic cut coverage to extend the computer geometry technique Cut Extension to propose Dynamic Cut Extension, which guides the camera toward an efficient path by calculating interaction values at different locations based on the visibility of multiple interactable objects.
\VRExplorer is a model-based testing tool to effectively interact with diverse virtual objects and explore complex VR scenes.
\XRintTest is an automated testing framework for Unity-based XR applications. It begins by constructing an XR User Interaction Graph, which models interaction targets and the events required to trigger them. Using this graph, the framework automatically explores the XR scene under test and generates user interactions
These five tools serve as baselines in our evaluation. 
To the best of our knowledge, they are the only available and operational approaches reported in the literature that support a fully \textit{automated} VR system testing. 

Following the statistical guidelines proposed by Arcuri et al.~\citep{arcuri2014hitchhiker}, each experiment was repeated 30~times to account for randomness of \VRGreed, \VRGuide, and \UltraInstinctVR.
Concerning \XRintTest and \VRExplorer, we evaluated each only once, as their approaches are deterministic and do not involve random algorithms.
To analyze and compare the performance of the different tools, we applied the non-parametric Wilcoxon signed-rank test with a significance level of $\alpha = 0.05$. In addition to statistical significance, we reported the Vargha--Delaney $\widehat{A}_{12}$ effect size~\citep{vargha_critique_2000} to quantify the magnitude of the observed differences between approaches.


\emph{Failures analysis (RQ2)}.
We map each failure to its corresponding source file using collected stack traces (\ie the current execution point when the failure occurred). Then we compute the frequency of each $<failure,location>$ pair for each tool. Finally, we identify unique failures based on their stack traces.

\emph{Defect identification (RQ3)}.
We select three non-functional failures that represent common issues in VR applications. Among these, we prioritize failures with the highest frequency, including one that consistently causes application crashes. In addition, we select one functional failure for further analysis.
Regarding the functional failure, we chose a case that occurs frequently across multiple projects and is representative of a requirement violation (\eg the inability to prevent movement through virtual objects such as a kettle).
We then perform a root cause analysis by examining execution logs and reproducing each failure based on the test scripts collected during the benchmarking process. Once reproduced, we isolate and analyze the failure to identify the underlying defect. This process involves inspecting the stack trace to locate the source file where the failure originates, and investigating potential causes using online resources such as developer forums (\eg Stack Overflow and Unity Discussions) to better understand and resolve the issue. After applying a fix, we classify the defect according to an established fault model, such as the one proposed by Andrade et al.~\citep{andrade2020understanding}.

\section{Results}

\subsection{Coverage (RQ1)}

\begin{figure}[t]
    \centering
    \includegraphics[width=\linewidth]{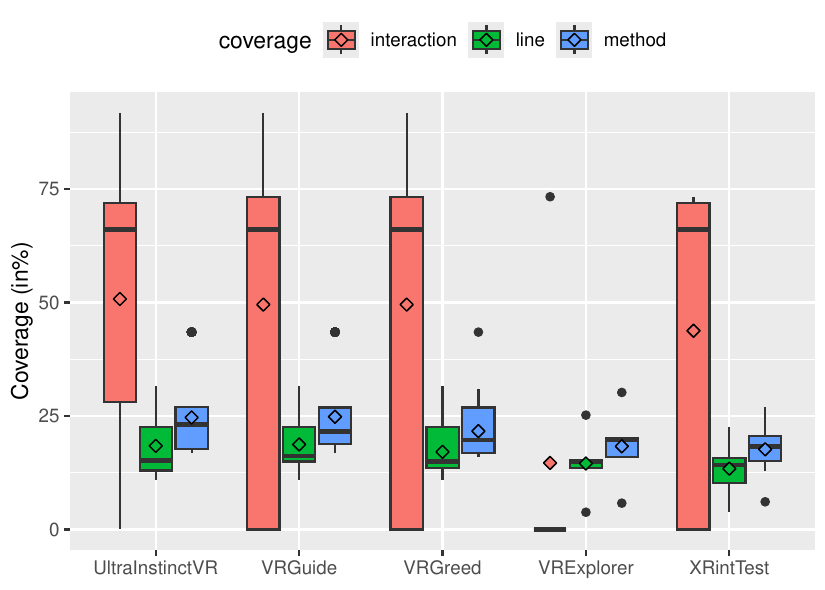}
    \caption{Coverage over 30, resp. 1, executions per benchmark for random-based (\UltraInstinctVR, \VRGuide, and \VRGreed), resp. exact (\VRExplorer and \XRintTest), approaches. Mean values are indicated by '$\diamond$'.}
    \label{fig:coverage}
\end{figure}


\Cref{fig:coverage} provides the distribution of each coverage metric across all evaluated projects. It is important to note that no coverage data could be retrieved for projects VertexForm3d, XRIExamples, and XRIStarterKit (also absent from Tables \ref{tab:benchmarkline} and \ref{tab:benchmarksinteraction}) due to early crashes during the application startup. This illustrates the difficulty of automating testing tasks for VR applications. 

Regarding \textbf{RQ1.1}, \UltraInstinctVR achieves a line coverage ranging from 12\% to 25\%. In comparison, \VRGuide attains between 15\% and 23\%, \VRGreed between 12\% and 23\%, \VRExplorer between 10\% and 12\%, and \XRintTest between 8\% and 15\%.

Regarding \textbf{RQ1.2}, \UltraInstinctVR achieves interaction coverage between 27\% and 72\%, whereas the other testing tools range from 0\% up to 75\%. Overall, these results indicate that \UltraInstinctVR performs comparably across coverage metrics while demonstrating stronger performance in specific cases.

\Cref{tab:benchmarkline} presents the statistical comparison between \UltraInstinctVR and the other methods. When comparing with \VRGuide, no definitive conclusion can be drawn except for \textit{EscapeRoom} as the effect sizes are negligible, suggesting that the observed differences may be due to randomness.
In contrast, when compared with \VRGreed, the results indicate that \UltraInstinctVR generally outperforms this approach, as evidenced by large effect sizes, with the exception of \textit{VRGameJamTemplate} and \textit{VRTemplate}, where no significant difference is observed.
Finally, with respect to \Cref{tab:benchmarksinteraction}, no clear conclusion can be drawn, as most effect sizes are negligible, indicating that the differences between approaches are not statistically significant. 

Some lines are reported as NA in the tables due to early application crashes during test suite execution, which prevented coverage data from being collected: XRIStarterAssets and WRToolkitEssential for \VRGuide and \VRGreed. Lines where no p-value could be computed (marked as NA) with a negligible magnitude are due to executions where the coverages were the same for all 30 executions for both \UltraInstinctVR and the other tool.

\begin{table*}[t]
    \centering
    \caption{Statistical comparison to \UltraInstinctVR for line coverage}
    \label{tab:benchmarkline}
    \footnotesize
    \begin{tabular}{lllllllllllll}
\toprule
\multicolumn{1}{c}{ } & \multicolumn{3}{c}{VRGuide} & \multicolumn{3}{c}{VRGreed} & \multicolumn{3}{c}{VRExplorer} & \multicolumn{3}{c}{XRintTest} \\
\cmidrule(l{3pt}r{3pt}){2-4} \cmidrule(l{3pt}r{3pt}){5-7} \cmidrule(l{3pt}r{3pt}){8-10} \cmidrule(l{3pt}r{3pt}){11-13}
\textbf{Benchmark} & \textbf{p-value} & \textbf{VD} & \textbf{Magnitude} & \textbf{p-value} & \textbf{VD} & \textbf{Magnitude} & \textbf{p-value} & \textbf{VD} & \textbf{Magnitude} & \textbf{p-value} & \textbf{VD} & \textbf{Magnitude}\\
\midrule
BAMConstruct & NA & 0.500 & negligible & \textbf{$<$0.001} & 0.017 & large & \textbf{$<$0.001} & $<$0.001 & large & \textbf{$<$0.001} & $<$0.001 & large\\
EscapeRoom & \textbf{$<$0.001} & 1.000 & large & \textbf{$<$0.001} & 1.000 & large & \textbf{$<$0.001} & 1.000 & large & \textbf{$<$0.001} & 1.000 & large\\
EscapeRoom-scene2 & 0.334 & 0.517 & negligible & \textbf{$<$0.001} & 0.049 & large & \textbf{$<$0.001} & 0.033 & large & \textbf{$<$0.001} & 0.033 & large\\
LayOffEvader & NA & 0.500 & negligible & \textbf{$<$0.001} & 0.017 & large & \textbf{$<$0.001} & $<$0.001 & large & \textbf{$<$0.001} & $<$0.001 & large\\
VRGameJamTemplate & NA & 0.500 & negligible & NA & 0.500 & negligible & \textbf{$<$0.001} & 1.000 & large & \textbf{$<$0.001} & 1.000 & large\\
\addlinespace
VRTemplate & NA & 0.500 & negligible & NA & 0.500 & negligible & \textbf{$<$0.001} & $<$0.001 & large & \textbf{$<$0.001} & $<$0.001 & large\\
XRIStarterAssets & NA & NA & NA & NA & NA & NA & \textbf{$<$0.001} & $<$0.001 & large & \textbf{$<$0.001} & $<$0.001 & large\\
XRToolkitEssantial & NA & NA & NA & NA & NA & NA & \textbf{$<$0.001} & $<$0.001 & large & \textbf{$<$0.001} & $<$0.001 & large\\
\bottomrule
\end{tabular}
\end{table*}

\begin{table*}[t]
    \centering
    \caption{Statistical comparison to \UltraInstinctVR for interaction coverage}
    \label{tab:benchmarksinteraction}
    \footnotesize
    \begin{tabular}{lllllllllllll}
\toprule
\multicolumn{1}{c}{ } & \multicolumn{3}{c}{VRGuide} & \multicolumn{3}{c}{VRGreed} & \multicolumn{3}{c}{VRExplorer} & \multicolumn{3}{c}{XRintTest} \\
\cmidrule(l{3pt}r{3pt}){2-4} \cmidrule(l{3pt}r{3pt}){5-7} \cmidrule(l{3pt}r{3pt}){8-10} \cmidrule(l{3pt}r{3pt}){11-13}
\textbf{Benchmark} & \textbf{p-value} & \textbf{VD} & \textbf{Magnitude} & \textbf{p-value} & \textbf{VD} & \textbf{Magnitude} & \textbf{p-value} & \textbf{VD} & \textbf{Magnitude} & \textbf{p-value} & \textbf{VD} & \textbf{Magnitude}\\
\midrule
BAMConstruct & NA & 0.500 & negligible & NA & 0.500 & negligible & NA & 0.500 & negligible & NA & 0.500 & negligible\\
EscapeRoom & NA & 0.500 & negligible & NA & 0.500 & negligible & \textbf{$<$0.001} & $<$0.001 & large & \textbf{$<$0.001} & $<$0.001 & large\\
EscapeRoom-scene2 & NA & 0.500 & negligible & NA & 0.500 & negligible & \textbf{$<$0.001} & $<$0.001 & large & \textbf{$<$0.001} & $<$0.001 & large\\
LayOffEvader & NA & 0.500 & negligible & NA & 0.500 & negligible & NA & 0.500 & negligible & NA & 0.500 & negligible\\
VRGameJamTemplate & NA & 0.500 & negligible & NA & 0.500 & negligible & \textbf{$<$0.001} & $<$0.001 & large & \textbf{$<$0.001} & $<$0.001 & large\\
\addlinespace
VRTemplate & NA & 0.500 & negligible & NA & 0.500 & negligible & \textbf{$<$0.001} & $<$0.001 & large & \textbf{$<$0.001} & $<$0.001 & large\\
XRIStarterAssets & NA & NA & NA & NA & NA & NA & \textbf{$<$0.001} & $<$0.001 & large & \textbf{$<$0.001} & $<$0.001 & large\\
XRToolkitEssentials & NA & NA & NA & NA & NA & NA & \textbf{$<$0.001} & $<$0.001 & large & \textbf{$<$0.001} & $<$0.001 & large\\
\bottomrule
\end{tabular}
\end{table*}

\begin{thebox}[title={RQ1. \UltraInstinctVR coverage efficiency}]
Regarding the effectiveness of \UltraInstinctVR in terms of code coverage, it is difficult to draw strong conclusions, as the statistical comparisons do not provide significant p-values to support our hypothesis. This limitation can be explained by the early crashes of the Unity application during the execution of the test suite, which reduce the amount of analyzable data and affect the reliability of the statistical results.
\end{thebox}

\subsection{Fault localization analysis (RQ2)}

\begin{table}
\addtolength{\tabcolsep}{-0.3em}
\caption{Unique failure analysis per tool with the number of failures common to the tool and \UltraInstinctVR (denoted $\cap$) and the number of failures detected only by the tool ($\setminus$)}
\label{tab:failure_by_method}
\small
\begin{tabular}{lccccc}
\toprule
Category & \scriptsize UltraInstinctVR & \scriptsize VRExplorer & \scriptsize VRGreed & \scriptsize VRGuide & \scriptsize XRIntTest \\
\midrule
Application crash & 21 & $\cap\,7 \;\setminus\,0$ & $\cap\,16 \;\setminus\,0$ & $\cap\,14 \;\setminus\,1$ & $\cap\,10 \;\setminus\,2$ \\
Dependency crash & 2 & $\cap\,1 \;\setminus\,0$ & $\cap\,1 \;\setminus\,0$ & $\cap\,1 \;\setminus\,0$ & $\cap\,1 \;\setminus\,0$ \\
Assertion violation & 7 & $\cap\,0 \;\setminus\,0$ & $\cap\,0 \;\setminus\,0$ & $\cap\,0 \;\setminus\,0$ & $\cap\,0 \;\setminus\,0$ \\
\midrule
\textbf{Unique failures} & \textbf{30} & \textbf{8} & \textbf{17} & \textbf{16} & \textbf{13} \\
\bottomrule
\end{tabular}
\end{table}

\Cref{tab:failure_by_method} shows that \VRGreed (17 unique failures) and \VRGuide (16 unique failures) detect largely overlapping sets of failures, indicating a high degree of similarity in the types of faults uncovered by these two approaches. In comparison, \VRExplorer detected 8 unique failures, while \XRintTest identified 13 unique failures, and \UltraInstinctVR detected 30 unique failures.
When examining overlaps with \UltraInstinctVR, 16 unique failures are shared with \VRGreed, 14 with \VRGuide, 8 with \VRExplorer, and 11 with \XRintTest. These results highlight that \UltraInstinctVR consistently detects a broader set of failures while still sharing a substantial subset of failures with each alternative approach.
Despite its overall coverage, \UltraInstinctVR failed to detect several unique failures identified by \XRintTest. Two of these failures come from \XRintTest tool itself, while another failure involves a component that should not have been activated but was erroneously triggered by an \XRintTest agent.

Across all evaluated projects, \UltraInstinctVR detected a total of 30 unique failures, substantially more than the unique failures identified by \VRGreed, \VRGuide, \XRintTest, and \VRExplorer. These results clearly demonstrate that \UltraInstinctVR is significantly more effective at uncovering unique failures during automated testing, largely due to its sophisticated oracle implementation. Furthermore, extending the number of implemented oracles is likely to increase the detection of additional unique failures.

\begin{thebox}[title={RQ2 Unique failure analysis}]
\UltraInstinctVR detects up to 30 unique failures, compared to \VRGreed which detected 16 unique failures, \VRGuide with 15 unique failures, \VRExplorer with 8 unique failures, and \XRintTest, identifying 13 unique failures. Since \UltraInstinctVR handles a broader range of faults than the other approaches, it is able to detect a greater number of failures.

\end{thebox}

\subsection{Root cause analysis (RQ3)}

\textbf{Functional root cause analysis}
The first failure we investigate concerns the “\textit{Gaze Interactor not set or missing}” error. This issue was triggered in the \textit{VRTemplate} project and detected by \UltraInstinctVR. It occurs when the gaze interactor is unavailable on the target device, as gaze-based interaction is primarily supported on Android devices and is not typically available on standard head-mounted displays (HMDs). We fixed this defect by disabling the \textit{XRGazeInteractor}. This failure is consistently detected by the other approaches part of the evaluation.

\begin{figure}[t]
    \includegraphics[width=88mm]{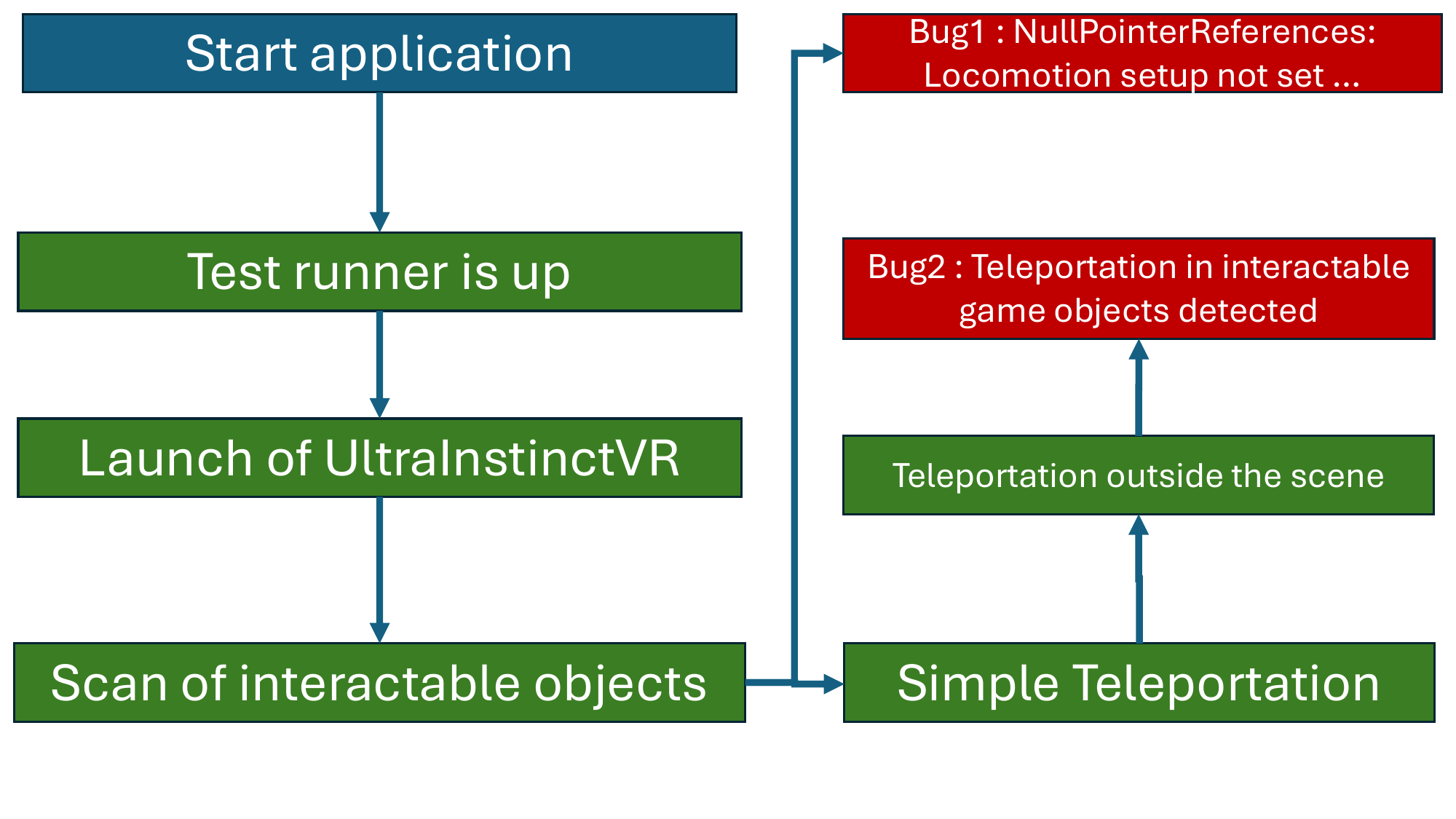}%
    \caption{Root cause analysis of VertexForm3D}
    \label{fig:rootcausefig}
\end{figure}

The second failure corresponds to an oracle failure implemented in \UltraInstinctVR, which occurs when an agent moves through a game object, as illustrated in \Cref{fig:rootcausefig} (Bug 2). To address this issue, we developed a script to prevent the player represented by a \textit{RigidBody} from passing through game objects (\eg scene walls or interactive objects).
This script is directly attached to the player, and we play on his behavior when interacting with virtual objects.
We were able to classify these two failures and their underlying defaults in the code in the fault model proposed by Andrade et al.~\citep{andrade2020understanding}.
The first default, related to the gaze interactor failure, is classified as a \textit{stability fault} (a crash). 
The second default corresponds to a problem with interacting with virtual objects.

\textbf{Non-functional root cause analysis.} 
\Cref{fig:rootcausefig} (Bug 1) illustrates the non-functional failure under investigation, which concerns \textit{NullReferenceException} errors. This issue mainly occurs in the \textit{VertexForm3D} project and is triggered at the start of the \UltraInstinctVR testing campaign, causing the application under test to crash. In particular, this failure is consistently detected by the other tools included in our benchmark.
This \textit{NullReferenceException} arises when two misconfigured rigid bodies collide, leading to an invalid object reference and ultimately resulting in a crash. To address this issue, we traced the defect back to the source file identified in the stack trace. Our analysis revealed that the failure is caused by a component assigned to \textit{null}, while its dependent component is missing from the application. To resolve the defect, the faulty component was disabled since this faulty component should not be present in the application, its deactivation does not remove any feature in the application.
According to the Andrade's fault model, this issue is classified as a \textit{stability fault} (a crash).

\begin{thebox}[title={RQ3 Real-world bug detection}]

\UltraInstinctVR can successfully detect real-world application bugs and, more importantly, is capable of identifying requirement-based failure.
\end{thebox}

\section{Threats to Validity}

Regarding \textit{external validity}, we follow a rigorous protocol to select projects that are representative of current VR development practices, taking into account factors such as community recognition and the use of widely adopted technologies.

Concerning \textit{internal validity}, our evaluation compares automated testing approaches based on the number of unique failures they detect. However, this metric may not fully reflect the actual number of underlying defects in the code. To mitigate this limitation, we complement our analysis by examining stack traces and analyzing the execution points of failures, as discussed in \textbf{RQ2}. An alternative approach would be to rely on a curated dataset of buggy programs, similar to Defects4J~\citep{just2014defects4j}; however, to the best of our knowledge, no such benchmark currently exists for VR applications.

Another threat to validity stems from the use of open-source VR projects, which are generally less mature than software in more established domains (e.g., Linux, Firefox). Furthermore, our evaluation does not include industrial VR applications, which typically exhibit higher levels of maturity in terms of maintenance practices, testing processes, and overall software quality.

Regarding the defects addressed in \textbf{RQ3}, although the applied fixes prevent \UltraInstinctVR from detecting the failures after correction, we cannot guarantee that these fixes do not introduce new defects. Therefore, the absence of detected failures should not be interpreted as definitive evidence of the correctness of the fixes.
\section{Discussion and future works}

Approaches such as \VRExplorer may exhibit poor code coverage performance. This can be explained by compatibility issues with the more recent Unity version used in our evaluation. In particular, \VRExplorer relies on native Unity components such as NavMesh, whose behavior and performance may vary across engine versions, potentially impacting the effectiveness of the approach.

Future work could better support handling exceptions to prevent the test campaign from crashing.
If this technical limitation applies to all the evaluated approaches, \UltraInstinctVR would increase its precision by overcoming it.



Regarding our approach, unlike more recent contributions such as \XRintTest and \VRExplorer, we did not focus on complex interaction generation, nor did we explore combinations of multiple interactions applied to the same virtual object. Instead, our work primarily emphasizes the detection of interaction-related failures and the development of mechanisms to define effective test oracles for VR applications

With respect to code coverage, software engineering practices for VR applications remain at an early stage of maturity. This may explain why, despite relatively low coverage percentages, a large number of unique failures can still be detected. These findings suggest that significant efforts are still needed to improve the automation of development and testing processes for VR applications, as previously highlighted by Andrade et al.~\citep{andrade2019towards}.

Additionally, for \textbf{RQ1}, we evaluate the effectiveness of \UltraInstinctVR in terms of code coverage. However, this raises an important question: to what extent is code coverage, as traditionally used in classical software engineering, a relevant metric for VR applications? Such applications heavily rely on external dependencies and engine-level components that are not directly implemented by developers but are still accounted for by coverage tools. Although it is possible to configure these tools to include or exclude such dependencies, code coverage may still fail to accurately reflect the quality of testing in this context.
Future work should therefore focus on defining more suitable evaluation metrics, as commonly explored in the HCI domain. These may include interaction coverage~\citep{lelli:hal-01123647} (\ie the extent to which virtual objects are interacted with), scene coverage, and coverage of the various parameters that govern the behavior of virtual objects and their associated interactions. 

Regarding defect localization, a current limitation of the evaluation of \UltraInstinctVR for \textbf{RQ2} is that, although it reports the file path where a fault originates, it does not always identify the specific method or code fragment responsible. Moreover, for failures detected through implemented oracles, the reported file does not necessarily correspond to the actual location of the underlying default to fix in the code, as it is the case with classical object-oriented testing approaches. More broadly, VR software testing is still in its nascent stage, mainly due to the absence of development standards and stable frameworks. In contrast to the mature, well-established
testing practices for web or desktop applications, current VR testing approaches remain exploratory and evolving. This situation reflects an early phase of methodological and tooling development, comparable to the state of software engineering several decades ago. Consequently, future work should focus on strengthening defect localization capabilities by adapting existing localization techniques to the specific characteristics
of VR systems.


\section{Conclusion}
Our research introduces \UltraInstinctVR, an automated framework for testing interaction-based behaviors in VR applications. The approach systematically generates and executes test scenarios targeting VR system interactions.
The experimental results show that \UltraInstinctVR is more effective at detecting unique failures than existing automated VR testing tools. In terms of code coverage, \UltraInstinctVR achieves comparable performance overall and, in several projects, demonstrates superior coverage, notably outperforming more recent tools such as \VRExplorer. Additionally, \UltraInstinctVR is capable of detecting real-world bugs. Future work may focus on better support of exceptions, novel coverage criteria dedicated to VR applications, and systematic approaches to help in precising the location of underlying defects based on a given failure.
Our results indicate that, despite relatively low coverage percentages particularly for line coverage—our approach is able to detect a large number of unique failures. This suggests that significant challenges remain in the automation of software development and testing for VR applications, highlighting the need for more advanced and tailored testing methodologies.


\subsection*{Data availability statement}
The replication package is publicly available and includes the \UltraInstinctVR tool, provided as \textit{UltraInstinctVR-ToolArtifact.zip}. The dataset and scripts used to produce the reported results are available in \textit{UltraInstinctVR-DataBenchmark-2984.zip}.
All data can be accessed via the Zenodo record at \url{https://zenodo.org/records/19248284}
with the associated DOI: \url{https://doi.org/10.5281/zenodo.19248284}.

\begin{acks}
This research was funded by the CyberExcellence (No. 2110186) and Wal4XR (No. 2310144) projects of DigitalWallonia, funded by the Public Service of Wallonia (SPW Recherche).
\end{acks}

\balance
\bibliographystyle{ACM-Reference-Format}
\bibliography{biblio}

@inproceedings{lemoulec:hal-01613873,
  TITLE = {{AGENT: Automatic Generation of Experimental Protocol Runtime}},
  AUTHOR = {Le Moulec, Gwendal and Argelaguet Sanz, Ferran and Gouranton, Val{\'e}rie and Blouin, Arnaud and Arnaldi, Bruno},
  URL = {https://hal.science/hal-01613873},
  BOOKTITLE = {{Virtual Reality Software and Technology}},
  ADDRESS = {Gothenburg, Sweden},
  SERIES = {Virtual Reality Software and Technology},
  YEAR = {2017},
  MONTH = Nov,
  HAL_ID = {hal-01613873},
  HAL_VERSION = {v1},
}

@article{10.1145/3715789,
author = {Cao, Shaoheng and Chen, Renyi and Yang, Wenhua and Pan, Minxue and Li, Xuandong},
title = {A Mixed-Methods Study of Model-Based GUI Testing in Real-World Industrial Settings},
year = {2025},
issue_date = {July 2025},
publisher = {Association for Computing Machinery},
address = {New York, NY, USA},
volume = {2},
number = {FSE},
url = {https://doi.org/10.1145/3715789},
doi = {10.1145/3715789},
journal = {Proc. ACM Softw. Eng.},
month = jun,
articleno = {FSE070},
numpages = {22},
}

@INPROCEEDINGS{6823874,
  author={Hunt, Chris J. and Brown, Guy and Fraser, Gordon},
  booktitle={2014 IEEE Seventh International Conference on Software Testing, Verification and Validation}, 
  title={Automatic Testing of Natural User Interfaces}, 
  year={2014},
  volume={},
  number={},
  pages={123-132},
  doi={10.1109/ICST.2014.25}}

@ARTICLE{6786194,
  author={Amalfitano, Domenico and Fasolino, Anna Rita and Tramontana, Porfirio and Ta, Bryan Dzung and Memon, Atif M.},
  journal={IEEE Software}, 
  title={MobiGUITAR: Automated Model-Based Testing of Mobile Apps}, 
  year={2015},
  volume={32},
  number={5},
  pages={53-59},
  doi={10.1109/MS.2014.55}}

@inproceedings{10.1145/2351676.2351717,
author = {Amalfitano, Domenico and Fasolino, Anna Rita and Tramontana, Porfirio and De Carmine, Salvatore and Memon, Atif M.},
title = {Using GUI ripping for automated testing of Android applications},
year = {2012},
publisher = {Association for Computing Machinery},
address = {New York, NY, USA},
doi = {10.1145/2351676.2351717},
booktitle = {Proceedings of the 27th IEEE/ACM International Conference on Automated Software Engineering},
pages = {258–261},
numpages = {4},
series = {ASE '12}
}

@inproceedings{lelli:hal-01123647,
  TITLE = {{On Model-Based Testing Advanced GUIs}},
  AUTHOR = {Lelli, Val{\'e}ria and Blouin, Arnaud and Baudry, Benoit and Coulon, Fabien},
  URL = {https://inria.hal.science/hal-01123647},
  BOOKTITLE = {{Software Testing, Verification and Validation Workshops (ICSTW), 2015 IEEE Eighth International Conference on}},
  ADDRESS = {Graz, Austria},
  PAGES = {1-10},
  YEAR = {2015},
  MONTH = Apr,
  DOI = {10.1109/ICSTW.2015.7107403},
}

@inproceedings{andrade2019towards,
  title={Towards the systematic testing of virtual reality programs},
  author={Andrade, Stev{\~a}o A and Nunes, Fatima LS and Delamaro, Marcio E},
  booktitle={2019 21st Symposium on Virtual and Augmented Reality (SVR)},
  pages={196--205},
  year={2019},
  organization={IEEE}
}

@inproceedings{lelli2015classifying,
  title={Classifying and qualifying GUI defects},
  author={Lelli, Val{\'e}ria and Blouin, Arnaud and Baudry, Benoit},
  booktitle={2015 IEEE 8th international conference on software testing, verification and validation (ICST)},
  pages={1--10},
  year={2015},
  organization={IEEE}
}

@inproceedings{rzig2023virtual,
  title={Virtual reality (vr) automated testing in the wild: A case study on Unity-based VR applications},
  author={Rzig, Dhia Elhaq and Iqbal, Nafees and Attisano, Isabella and Qin, Xue and Hassan, Foyzul},
  booktitle={Proceedings of the 32nd ACM SIGSOFT International Symposium on Software Testing and Analysis},
  pages={1269--1281},
  year={2023}
}

@inproceedings{ashtari_creating_2020,
	address = {Honolulu HI USA},
	title = {Creating {Augmented} and {Virtual} {Reality} {Applications}: {Current} {Practices}, {Challenges}, and {Opportunities}},
	shorttitle = {Creating {Augmented} and {Virtual} {Reality} {Applications}},
	doi = {10.1145/3313831.3376722},
	booktitle = {Proceedings of the 2020 {CHI} {Conference} on {Human} {Factors} in {Computing} {Systems}},
	publisher = {ACM},
	author = {Ashtari, Narges and Bunt, Andrea and McGrenere, Joanna and Nebeling, Michael and Chilana, Parmit K.},
	month = apr,
	year = {2020},
	pages = {1--13},
}

@inproceedings{10.1145/3660515.3664244,
author = {Septon, Thibaut and Villarreal-Narvaez, Santiago and Devroey, Xavier and Dumas, Bruno},
title = {Exploiting Semantic Search and Object-Oriented Programming to Ease Multimodal Interface Development},
year = {2024},
publisher = {ACM},
doi = {10.1145/3660515.3664244},
booktitle = {Companion Proceedings of the 16th ACM SIGCHI Symposium on Engineering Interactive Computing Systems},
pages = {74–80},
numpages = {7},
location = {Cagliari, Italy},
series = {EICS '24 Companion}
}

@inproceedings{gu2019practical,
  title={Practical GUI testing of Android applications via model abstraction and refinement},
  author={Gu, Tianxiao and Sun, Chengnian and Ma, Xiaoxing and Cao, Chun and Xu, Chang and Yao, Yuan and Zhang, Qirun and Lu, Jian and Su, Zhendong},
  booktitle={2019 IEEE/ACM 41st International Conference on Software Engineering (ICSE)},
  pages={269--280},
  year={2019},
  organization={IEEE}
}

@inproceedings{wang2022vrtest,
  title={Vrtest: An extensible framework for automatic testing of virtual reality scenes},
  author={Wang, Xiaoyin},
  booktitle={Proceedings of the ACM/IEEE 44th International Conference on Software Engineering: Companion Proceedings},
  pages={232--236},
  year={2022}
}

@inproceedings{andrade2020understanding,
  title={Understanding VR software testing needs from stakeholders’ points of view},
  author={Andrade, Stev{\~a}o A and Quevedo, Alvaro Joffre U and Nunes, Fatima LS and Delamaro, M{\'a}rcio E},
  booktitle={2020 22nd Symposium on Virtual and Augmented Reality (SVR)},
  pages={57--66},
  year={2020},
  organization={IEEE}
}

@article{arcuri2014hitchhiker,
  title={A hitchhiker's guide to statistical tests for assessing randomized algorithms in software engineering},
  author={Arcuri, Andrea and Briand, Lionel},
  journal={Software Testing, Verification and Reliability},
  volume={24},
  number={3},
  pages={219--250},
  year={2014},
  publisher={Wiley Online Library}
}

@inproceedings{claude2014short,
  title={Short paper:\# seven, a sensor effector based scenarios model for driving collaborative virtual environment},
  author={Claude, Guillaume and Gouranton, Val{\'e}rie and Berthelot, Rozenn Bouville and Arnaldi, Bruno},
  booktitle={ICAT-EGVE, International Conference on Artificial Reality and Telexistence, Eurographics Symposium on Virtual Environments},
  pages={1--4},
  year={2014}
}

@inproceedings{wang2023vrguide,
  title={VRGuide: Efficient Testing of Virtual Reality Scenes via Dynamic Cut Coverage},
  author={Wang, Xiaoyin and Rafi, Tahmid and Meng, Na},
  booktitle={2023 38th IEEE/ACM International Conference on Automated Software Engineering (ASE)},
  pages={951--962},
  year={2023},
  organization={IEEE}
}

@misc{Unity,
  author = {Unity},
  title = {Unity Technologies},
  howpublished = {\url{https://learn.unity.com/}},
  month = {February},
  year = {2025},
}

@article{correa2021software,
  title={Software testing automation of VR-based systems with haptic interfaces},
  author={Corr{\^e}a, Cl{\'e}ber G and Delamaro, M{\'a}rcio E and Chaim, Marcos L and Nunes, F{\'a}tima LS},
  journal={The Computer Journal},
  volume={64},
  number={5},
  pages={826--841},
  year={2021},
  publisher={Oxford University Press}
}

@misc{Xareus,
  author = {Gouranton, Valérie and Nouviale Florian},
  title = { Xareus
A set of tools to ease the development of XR applications.  },
  howpublished = {\url{https://xareus.insa-rennes.fr}},
  month = {February},
  year = {2025},
}

@misc{GithubAPI,
  author = {Github},
  title = {GitHub REST API documentation},
  howpublished = {\url{https://docs.github.com/en/rest?apiVersion=2022-11-28}},
  month = {September},
  year = {2025},
}

@inproceedings{sutherland1965ultimate,
  title={The ultimate display},
  author={Sutherland, Ivan E and others},
  booktitle={Proceedings of the IFIP Congress},
  volume={2},
  number={506-508},
  pages={506--508},
  year={1965},
  organization={New York}
}

@article{agrawal2019defining,
  title={Defining immersion: Literature review and implications for research on immersive audiovisual experiences},
  author={Agrawal, Sarvesh and Simon, Ad{\`e}le and Bech, S{\o}ren and B{\ae}rentsen, Klaus and Forchhammer, S{\o}ren},
  journal={Journal of Audio Engineering Society},
  volume={68},
  number={6},
  pages={404--417},
  year={2019},
  publisher={Audio Engineering Society}
}

@inproceedings{lee2023comparison,
  title={Comparison of virtual reality teleportation targeting method performance depending on the teleport distance},
  author={Lee, Jihyeon and Kim, Jinwook and Lee, Jeongmi},
  booktitle={2023 IEEE International Symposium on Mixed and Augmented Reality Adjunct (ISMAR-Adjunct)},
  pages={742--745},
  year={2023},
  organization={IEEE}
}

@inproceedings{atienza2016interaction,
  title={Interaction techniques using head gaze for virtual reality},
  author={Atienza, Rowel and Blonna, Ryan and Saludares, Maria Isabel and Casimiro, Joel and Fuentes, Vivencio},
  booktitle={2016 IEEE Region 10 Symposium (TENSYMP)},
  pages={110--114},
  year={2016},
  organization={IEEE}
}

@ARTICLE{10458415,
  author={Javerliat, Charles and Villenave, Sophie and Raimbaud, Pierre and Lavoué, Guillaume},
  journal={IEEE Transactions on Visualization and Computer Graphics}, 
  title={PLUME: Record, Replay, Analyze and Share User Behavior in 6DoF XR Experiences}, 
  year={2024},
  volume={30},
  number={5},
  pages={2087-2097},
  doi={10.1109/TVCG.2024.3372107}}

@inproceedings{gramoli2025xareus,
  title={Xareus: a Framework to Create Interactive Applications without Coding},
  author={Gramoli, Lysa and Nouviale, Florian and Reuzeau, Adrien and Audinot, Alexandre and Risy, Mathieu and Marchand-Guerniou, Tangui and Mavromatis, Ma{\'e} and Arnaldi, Bruno and Gouranton, Val{\'e}rie},
  booktitle={2025 IEEE Conference on Virtual Reality and 3D User Interfaces Abstracts and Workshops (VRW)},
  pages={1658--1659},
  year={2025},
  organization={IEEE}
}

@book{utting_practical_2007,
    title = {Practical {Model}-{Based} {Testing}: {A} {Tools} {Approach}},
    publisher = {Morgan Kaufmann},
    author = {Utting, Mark and Legeard, Bruno},
    year = {2007},
}

@article{spittle2022review,
  title={A review of interaction techniques for immersive environments},
  author={Spittle, Becky and Frutos-Pascual, Maite and Creed, Chris and Williams, Ian},
  journal={IEEE Transactions on Visualization and Computer Graphics},
  volume={29},
  number={9},
  pages={3900--3921},
  year={2022},
  publisher={IEEE}
}

@article{gu2025software,
  title={Software testing for extended reality applications: a systematic mapping study},
  author={Gu, Ruizhen and Rojas, Jos{\'e} Miguel and Shin, Donghwan},
  journal={Automated Software Engineering},
  volume={32},
  number={2},
  pages={56},
  year={2025},
  publisher={Springer}
}

@article{vargha_critique_2000,
	title = {A {Critique} and {Improvement} of the {CL} {Common} {Language} {Effect} {Size} {Statistics} of {McGraw} and {Wong}},
	volume = {25},
	doi = {10.3102/10769986025002101},
	number = {2},
	journal = {Journal of Educational and Behavioral Statistics},
	author = {Vargha, András and Delaney, Harold D.},
	month = jun,
	year = {2000},
	pages = {101--132},
}

@article{parry2021survey,
  title={A survey of flaky tests},
  author={Parry, Owain and Kapfhammer, Gregory M and Hilton, Michael and McMinn, Phil},
  journal={ACM Transactions on Software Engineering and Methodology (TOSEM)},
  volume={31},
  number={1},
  pages={1--74},
  year={2021},
  publisher={ACM New York, NY}
}

@article{calvary2014introduction,
  title={Introduction to model-based user interfaces},
  author={Calvary, Ga{\"e}lle and Coutaz, Jo{\"e}lle},
  journal={Group Note},
  volume={7},
  pages={W3C},
  year={2014}
}

@book{aniche2022effective,
  title={Effective Software Testing: A developer's guide},
  author={Aniche, Maur{\'\i}cio},
  year={2022},
  publisher={Simon and Schuster}
}

@inproceedings{prasetya2020navigation,
  title={Navigation and exploration in 3D-game automated play testing},
  author={Prasetya, ISWB and Voshol, Maurin and Tanis, Tom and Smits, Adam and Smit, Bram and Mourik, Jacco van and Klunder, Menno and Hoogmoed, Frank and Hinlopen, Stijn and Casteren, August van and others},
  booktitle={Proceedings of the 11th ACM SIGSOFT International Workshop on Automating TEST Case Design, Selection, and Evaluation},
  pages={3--9},
  year={2020}
}

@inproceedings{ghrairi2018state,
  title={The state of practice on virtual reality (vr) applications: An exploratory study on github and stack overflow},
  author={Ghrairi, Naoures and Kpodjedo, S{\`e}gla and Barrak, Amine and Petrillo, Fabio and Khomh, Foutse},
  booktitle={2018 IEEE International Conference on Software Quality, Reliability and Security (QRS)},
  pages={356--366},
  year={2018},
  organization={IEEE}
}

@inproceedings{prasetya2021agent,
  title={An agent-based architecture for ai-enhanced automated testing for xr systems, a short paper},
  author={Prasetya, ISWB and Shirzadehhajimahmood, Samira and Ansari, Saba Gholizadeh and Fernandes, Pedro and Prada, Rui},
  booktitle={2021 IEEE International Conference on Software Testing, Verification and Validation Workshops (ICSTW)},
  pages={213--217},
  year={2021},
  organization={IEEE}
}

@inproceedings{just2014defects4j,
  title={Defects4J: A database of existing faults to enable controlled testing studies for Java programs},
  author={Just, Ren{\'e} and Jalali, Darioush and Ernst, Michael D},
  booktitle={Proceedings of the 2014 international symposium on software testing and analysis},
  pages={437--440},
  year={2014}
}

@inproceedings{gu2025test,
  title={A test automation framework for user interaction in extended reality applications},
  author={Gu, Ruizhen and Rojas, Jos{\'e} Miguel},
  booktitle={2025 40th IEEE/ACM International Conference on Automated Software Engineering Workshops (ASEW)},
  pages={325--330},
  year={2025},
  organization={IEEE}
}

@inproceedings{gu2025xrinttest,
  title={XRintTest: An automated framework for user interaction testing in extended reality applications},
  author={Gu, Ruizhen and Rojas, Jos{\'e} Miguel and Shin, Donghwan},
  booktitle={2025 40th IEEE/ACM International Conference on Automated Software Engineering (ASE)},
  pages={4013--4016},
  year={2025},
  organization={IEEE}
}

@inproceedings{zhu2025vrexplorer,
  title={VRExplorer: A Model-based Approach for Semi-Automated Testing of Virtual Reality Scenes},
  author={Zhu, Zhengyang and Dai, Hong-Ning and Guo, Hanyang and Liao, Zeqin and Zheng, Zibin},
  booktitle={2025 40th IEEE/ACM International Conference on Automated Software Engineering (ASE)},
  pages={482--494},
  year={2025},
  organization={IEEE}
}
\end{document}